# Confluently Persistent Sets and Maps

Olle Liljenzin

**Abstract**
Ordered sets and maps play important roles as index structures in relational data models. When a shared index in a multi-user system is modified concurrently, the current state of the index will diverge into multiple versions containing the local modifications performed in each work flow. The confluent persistence problem arises when versions should be melded in commit and refresh operations so that modifications performed by different users become merged.

Confluently Persistent Sets and Maps are functional binary search trees that support efficient set operations both when operands are disjoint and when they are overlapping. Treap properties with hash values as priorities are maintained and with hash-consing of nodes a unique representation is provided. Non-destructive set merge algorithms that skip inspection of equal subtrees and a conflict detecting meld algorithm based on set merges are presented. The meld algorithm is used in commit and refresh operations. With m modifications in one flow and n items in total, the expected cost of the operations is O(m log(n/m)).

## Introduction

A data structure is a unique representation of a data set if all occurrences of the same data set are represented by a single copy of the structure. Provided unique representation equality of data sets can be deduced by address comparison at a cost of O(1). A balanced search tree will be uniquely represented if all nodes are uniquely represented and the balancing algorithm is deterministic and independent from history so that the shape of a tree is a pure function of the current data set.

Hash-consing is a technique used to avoid duplication of structurally equal objects. Created objects are stored in hash tables and equal objects are reused before new objects are created [1].

Ordinary data structures that provide access to the current version only are called ephemeral. A data structure is called persistent if it can be modified and previous versions can be accessed after the modification [2]. A method is called non-destructive if it modifies a data structure without use of assignments so that the old version of the structure is still available after the modification. A data structure is called functional if all methods used to modify it are non-destructive. A functional data structure is automatically persistent [3].

A persistent data structure is created in an initial version and each modification of it will create a new version. If always the newest version is modified a time line of historical versions is formed. If any version other than the newest is modified, the time line of versions will be branched into a version tree and branches will represent concurrent flows. A data structure is called confluently persistent if there is a meld operation that creates a new version from two previous versions so that branches in a version tree are joined and a version DAG is formed [4].

In a system that allows concurrent modifications of a shared data set, the problem of confluent persistence arises when modifications should be merged. Concurrent modifications will create branched flows of new versions and the current version in each flow will represent a local view of data. Modifications performed by one user will not automatically be visible to other users and to make them visible operations that synchronize flows are required. Commit operations update a shared data set with local modifications and refresh operations update a local flow with modifications from a shared data set. Application level modifications will often consist of multiple atomic modifications forming transactions that must be performed entirely or be aborted with no effect. Commit and refresh are meld operations that create a new version from previous versions.

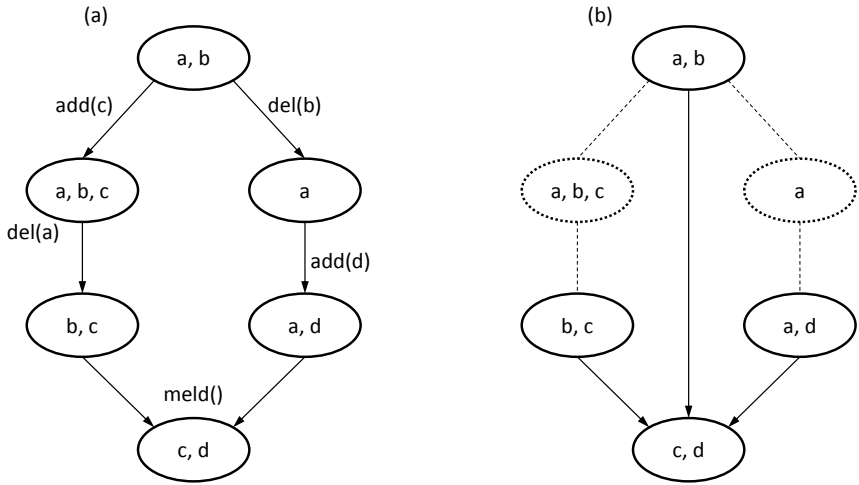

**Figure 1. Meld operations on sets**

Figure 1a shows versions of a set created by modifications in two concurrent flows followed by a meld operation that joins the flows with the result of the least common ancestor modified sequentially. The accumulated modifications in each branch are the difference between the melded version and the least common ancestor. With **meld()** defined as a function of the current content in three given versions, as shown in Figure 1b, access to intermediate versions is not required. In repeated refresh and commit operations the ancestor will be replaced by a more recent version to prevent the same modifications from being merged more than once.

A meld is valid if the modifications in both version branches can be performed sequentially, otherwise it is invalid. Figure 2 shows invalid melds as a result of conflicting modifications. A meld operation is conflict detecting if invalid melds are aborted.

Maps are sets of keys and associations and we meld maps by melding the sets. A meld of maps is valid if the meld of contained keys is valid and the meld of contained associations is valid.

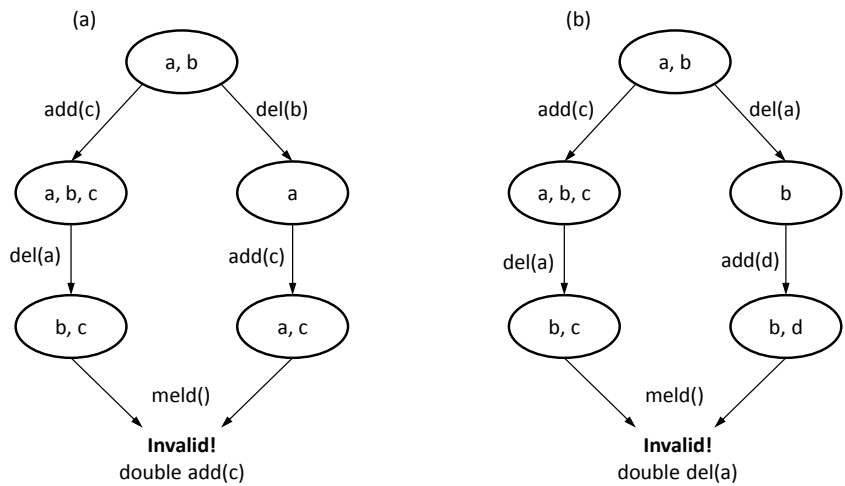

**Figure 2. Invalid melds**

Figure 3 shows **refresh** and **commit** operations in a multi-user system. A local flow is associated with an ancestor version and a current local version. The initial ancestor is the branched version. Refresh and commit operations are melds of current local version and current shared version. In case of **refresh** the result of the meld becomes new local version and the current shared version becomes new ancestor. In case of **commit** the result of the meld becomes new shared version and the current local version becomes new ancestor.

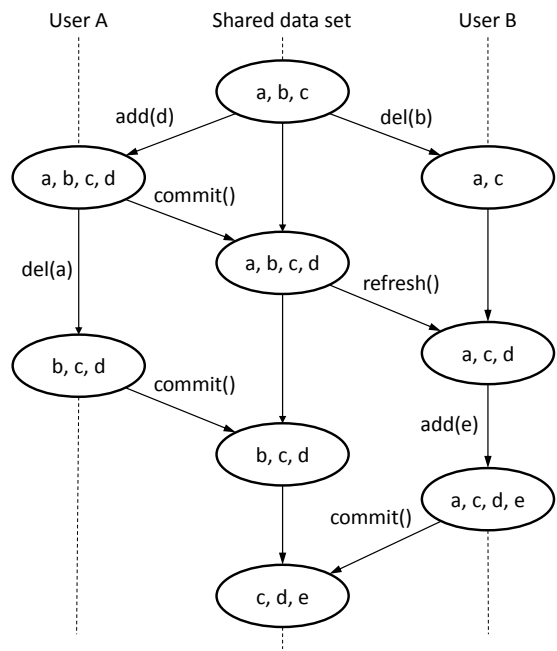

**Figure 3. Refresh and commit operations in a multi-user system**

## Related work

Driscoll et al. addressed the problem of making a general pointer based data structure persistent [2] and defined the special case of confluent persistence [4]. Confluently persistent solutions have been presented for some structures such as catenable lists [4], deques [5] and tries [6].

Fiat and Kaplan [7] followed by Collette, Iacono and Langerman [8] presented transformations for making general pointer based data structures confluently persistent. They addressed persistent representation, access paths and cost of using them.

Treaps were introduced by Seidel and Aragon in Randomized Search Trees [9]. Treaps are binary search trees with probabilistic balancing. A node contains a key and references to left and right child nodes, which might be null. Nodes are assigned random priorities. Key order is maintained so the contained key splits subtrees into smaller values in the left child and greater values in the right child. Heap order is maintained so that the priority of a parent node is greater than the priority of any child node. The expected height of a Treap is asymptotically logarithmic, i.e. E[h] is O(log n). The authors suggested the random number generator could be replaced by a general hash function to solve the unique representation problem.

Blelloch and Reid-Miller [10] presented set operations on Treaps requiring an expected cost of O(m log(n/m)) for merging a larger set of size n with a smaller set of size m. They used divide and conquer algorithms that split both input sets by the key with top priority always found in a root node. The split key is excluded in recursive calls, thus the height of at least one input tree is reduced in each recursion.

## Algorithms

Python code is used to describe algorithms.

```
class Node(object):
    def __init__(self, key, left, right):
        self.key = key; self.left = left; self.right = right

    def __eq__(self, other):
        return self.left is other.left and self.right is other.right \
            and self.key == other.key

    def __hash__(self):
        return hash((self.key, id(self.left), id(self.right)))
```

The empty Treap is represented by a null reference (**None** in Python). Other Treaps are represented by references to root nodes. A Node has a key and references to left and right children.

```
import weakref

nodes = {}

def Treap(key, left = None, right = None):
    node = Node(key, left, right); ref = weakref.ref(node)
    if ref not in nodes:
        nodes[weakref.ref(node, lambda x : nodes.pop(x))] = ref
    return nodes[ref]()
```

Treaps are unique if nodes are unique, thus created nodes are stored in hash maps (dictionaries in Python) and existing nodes are reused before new nodes are created. The function **Treap()** is a node factory always returning unique nodes when called. Created nodes should be considered immutable and must not be modified.

```
import hashlib

def priority(t):
    return hashlib.sha512(repr(t.key).encode()).digest()
```

It is assumed the string representation of keys can be generated at a cost of O(1) and that keys are equal if their string representations are equal. A cryptographically safe 512-bit hash function is used to avoid discussions on quality of hash functions in this context.

```
def join(t1, t2):
    if t1 is None: return t2
    if t2 is None: return t1
    if priority(t1) < priority(t2):
        return Treap(t2.key, join(t1, t2.left), t2.right)
    else:
        return Treap(t1.key, t1.left, join(t1.right, t2))
```

The function **join()** recurses down in the source trees and creates the result on the way up after the recursive call. The cost is bounded by the height of the result which is at most T(2h).

```
def split(t, key):
    if t is None: return (None, None)
    if key < t.key:
        l1, l2 = split(t.left, key)
        return (l1, Treap(t.key, l2, t.right))
    else:
        r1, r2 = split(t.right, key)
        return (Treap(t.key, t.left, r1), r2)
```

The function **split()** recurses down in the source tree and creates the result on the way up after the recursive call. The cost is bounded by the height of the operand which is at most T(h).

```
def union(t1, t2):
    if t1 is t2 or t2 is None: return t1
    if t1 is None: return t2
    if t1.key == t2.key:
        return Treap(t1.key, union(t1.left, t2.left), union(t1.right, t2.right))
    elif priority(t1) < priority(t2):
        l1, r1 = split(t1, t2.key)
        return Treap(t2.key, union(l1, t2.left), union(r1, t2.right))
    else:
        l2, r2 = split(t2, t1.key)
        return Treap(t1.key, union(t1.left, l2), union(t1.right, r2))
```

The function **union()** merges t1 and t2 and the resulting Treap will contain all keys in any of t1 or t2.

```
def intersection(t1, t2):
    if t1 is t2: return t1
    if t1 is None or t2 is None: return None
    if t1.key == t2.key:
        return Treap(t1.key, intersection(t1.left, t2.left), \
            intersection(t1.right, t2.right))
    elif priority(t1) < priority(t2):
        l1, r1 = split(t1, t2.key)
        return join(intersection(l1, t2.left), intersection(r1, t2.right))
    else:
        l2, r2 = split(t2, t1.key)
        return join(intersection(t1.left, l2), intersection(t1.right, r2))
```

The function **intersection()** merges t1 and t2 and the resulting Treap will contain all keys in both t1 and t2.

```
def difference(t1, t2):
    if t1 is t2 or t1 is None: return None
    if t2 is None: return t1
    if t1.key == t2.key:
        return join(difference(t1.left, t2.left), difference(t1.right, t2.right))
    elif priority(t1) < priority(t2):
        l1, r1 = split(t1, t2.key)
        return join(difference(l1, t2.left), difference(r1, t2.right))
    else:
        l2, r2 = split(t2, t1.key)
        return Treap(t1.key, difference(t1.left, l2), difference(t1.right, r2))
```

The function **difference()** merges t1 and t2 and the resulting Treap will contain all keys in t1 but not in t2.

```
def symmetric_difference(t1, t2):
    if t1 is t2: return None
    if t1 is None: return t2
    if t2 is None: return t1
    if t1.key == t2.key:
        return join(symmetric_difference(t1.left, t2.left), \
            symmetric_difference(t1.right, t2.right))
    elif priority(t1) < priority(t2):
        l1, r1 = split(t1, t2.key)
        return Treap(t2.key, symmetric_difference(l1, t2.left), \
            symmetric_difference(r1, t2.right))
    else:
        l2, r2 = split(t2, t1.key)
        return Treap(t1.key, symmetric_difference(t1.left, l2), \
            symmetric_difference(t1.right, r2))
```

The function **symmetric_difference()** merges t1 and t2 and the resulting Treap will contain all keys in exactly one of t1 or t2.

Blelloch and Reid-Miller [10] proved an expected cost of O(m log(n/m)) for their algorithms, where n is the size of the larger operand and m is the size of the smaller operand. The algorithms presented here differ from the algorithms presented in [10] as a conditional exit is introduced; "if t1 is t2: return *t1, t2 or None*". Operands are compared at a cost of T(1) with a possible outcome of earlier exit. The added test cannot increase complexity and the algorithms presented here are also bounded by O(m log(n/m)).

The recursive call tree is same in all four set operations. It follows the pattern of *merge*():

```
def merge(t1, t2):
    if t1 is t2 or t1 is None or t2 is None:
        if some condition: return t1
        elif some other condition: return t2
        else: return None
    key = the key with top priority
    l1, r1 = split(t1, key); l2, r2 = split(t2, key)
    if some condition: return Treap(key, merge(l1, r1), merge(l2, r2))
    else: return join(merge(l1, r1), merge(l2, r2))
```

Let n be max(|t1|, |t2|), let m be min(|t1|, |t2|), let d be |t1 Δ t2|, let k be min (d, m) and let G be the recursive call tree. Let S and S' be t1 and t2 with priority(S) ≥ priority(S'), let S" be t1 Δ t2 and let h be the height of S.

We shall prove *merge*() has an expected cost of O(k log(n/k)).

The split key is selected from S and recursion stops if any of S' or S" is ∅. Branches in G are created as a function of S and pruned as a function of S' and S". The internal cost of T(h) + T(2h) for calls to split() and join() is not exceeding the stated cost of O(k log(n/k)). Thus merge() must be O(k log(n/k)) because it is O(m log(n/m)) and substituting S' by S" will not change G.

Thus the following bound for set operations:

$$\left.\begin{array}{l}A \cup B \\ A \cap B \\ A \setminus B \\ A \Delta B\end{array}\right\} \in O\left(\min(|A|,|B|,|A\Delta B|) \log\left(\frac{|A|+|B|}{\min(|A|,|B|,|A\Delta B|)}\right)\right) \quad expected\ cost$$

```
def meld(t0, t1, t2):
    d1 = difference(t0, t1)
    a1 = difference(t1, t0)
    d2 = difference(t0, t2)
    a2 = difference(t2, t0)
    dd = intersection(d1, d2)
    aa = intersection(a1, a2)
    if dd is None and aa is None:
        return union(difference(t2, d1), a1)
    else:
        raise KeyError
```

The function **meld()** merges modifications from t0 to t1 and modifications from t0 to t2. KeyError is raised in case of duplicated add or delete modifications. Let n be the size of the largest set t0, t1 and t2 and let m be the size of the largest set t0 Δ t1, t0 Δ t2 and t1 Δ t2. The cost of meld() is the cost of the merge operations, thus O(m log(n/m)) expected cost.

## Implementation details

Approximately half the work in meld() is to find out if dd and aa are empty and the meld is valid. By specialized merge functions we could move the conflict detection inside the final merges to reduce the number of merges.

```
def add(t1, t2):
    if t1 is None: return t2
    if t2 is None: return t1
    if t1.key == t2.key: raise KeyError
    if priority(t1) < priority(t2):
        l1, r1 = split(t1, t2.key)
        return Treap(t2.key, add(l1, t2.left), add(r1, t2.right))
    else:
        l2, r2 = split(t2, t1.key)
        return Treap(t1.key, add(t1.left, l2), add(t1.right, r2))
```

The function **add()** merges t1 and t2 with set union semantics, but raises KeyError if the intersection is not empty.

```
def sub(t1, t2):
    if t1 is t2: return None
    if t1 is None: raise KeyError
    if t2 is None: return t1
    if t1.key == t2.key:
        return join(sub(t1.left, t2.left), sub(t1.right, t2.right))
    if priority(t1) < priority(t2):
        raise KeyError
    l2, r2 = split(t2, t1.key)
    return Treap(t1.key, sub(t1.left, l2), sub(t1.right, r2))
```

The function **sub()** merges t1 and t2 with set difference semantics, but raises KeyError if t2 is not a subset of t1.

```
def meld(t0, t1, t2):
    return add(sub(t2, difference(t0, t1)), difference(t1, t0))
```

The function **meld()** is reduced to four merges. The cost is O(m log(n/m)) with m = |t0 Δ t1|. Thus if we have reasons to expect |t0 Δ t2| < |t0 Δ t1| we should swap the order of t1 and t2 to optimize performance.

When implementing maps they will be sets of key-value pairs. The priority is the hash value of the key. We need two versions of the algorithms; one version that compares keys and a second version that compares keys first and then values if keys are equal. The first version should be used when updating maps with new key-value pairs to maintain uniqueness of keys. The second version is the one that creates the new state, and also detects value conflicts during meld. When melding maps we must run both versions of the meld algorithm to detect both sorts of conflicts and if no conflicts were found we will return the result of the second version.

For applications requiring fast access to aggregate values we could store the result in nodes as we create them. The following code example provides fast access to cardinality:

```python
def cardinality(t):
    if t is None: return 0
    return t.cardinality

class Node(object):
    def __init__(self, key, left, right):
        self.key = key; self.left = left; self.right = right
        self.cardinality = 1 + cardinality(left) + cardinality(right)

    def __eq__(self, other):
        return self.left is other.left and self.right is other.right \
            and self.key == other.key

    def __hash__(self):
        return hash((self.key, id(self.left), id(self.right)))
```

In highly tuned applications the cost of using cryptographic hash functions might be inappropriate. As an alternative we suggest CRC-32C [11] for randomizing the result of ordinary hash functions with low collision probability. In recent processors implementing the SSE 4.2 instruction set, hardware support is provided for CRC-32C and the cost of using it is that of a machine instruction:

```cpp
template <class Key>
struct Treap {
    std::atomic<size_t> refcount;
    Treap *left, *right;
    Key key;
};

template <class Key>
size_t priority(const Treap<Key> &treap) {
    return __builtin_ia32_crc32di(0, std::hash<Key>()(treap.key));
}
```

The example code in C++ with macros for the GNU g++ compiler is useful for sets containing up to $2^{32}$ keys. Note the CRC-32C function will not be collision free and when comparing priorities the keys should be compared if hash values are equal.

If the node factory is guarded the presented algorithms could be parallelized as in [10].

## Examples

The following program produces the flow shown in figure 3:

```
# Create shared data set S with keys { 'a', 'b', 'c' }
S = None
for key in ('a', 'b', 'c'):
    S = add(S, Treap(key))

# Create local flow A by adding 'd' to shared data set
# A[0] is current local version and A[1] is ancestor version
A = [ add(S, Treap('d')), S ]

# Create local flow B by deleting 'b' from shared data set
# B[0] is current local version and B[1] is ancestor version
B = [ sub(S, Treap('b')), S ]

# Commit modifications in A
S = meld(A[1], A[0], S); A[1] = A[0]

# Delete 'a' from A
A[0] = sub(A[0], Treap('a'))

# Refresh B
B = [ meld(B[1], B[0], S), S ]

# Commit modifications in A
S = meld(A[1], A[0], S); A[1] = A[0]

# Add 'e' to B
B[0] = add(B[0], Treap('e'))

# Commit modifications in B
S = meld(B[1], B[0], S); B[1] = B[0]

# Define a generator so that we can iterate over keys in a set
def iterate(t):
    if t is not None:
        for k in iterate(t.left): yield k
        yield t.key
        for k in iterate(t.right): yield k

# Print the current version of S
print('S = { %s }' % ', '.join(map(repr, iterate(S))))
```

The arguments to meld are ordered to be optimal in case of local modifications being less frequent than modifications of the shared data set.

The output of the program is the shared data set after the commit operations:

```
S = { 'c', 'd', 'e' }
```

## Conclusions

Unique representation provides a clean solution to the confluent persistence problem for ordered sets and maps. Using hashed priorities the structure of Treaps is based on the current content only, thus the representation of a version is always optimal and cannot age when new versions are added in a system. There is no structural dependency between versions and exclusively used memory is released when versions are no longer accessible.

Each Treap is an ephemeral tree and access performance is that of balanced search trees. Path copying and hash consing of nodes add to update times, while functional data structures provide a stable platform for parallel execution.

A meld operation is a series of merges of two current versions and an ancestor version. The result is intuitive both when melds are valid and when yielding conflicts. Refresh and commit operations are melds that move the ancestor version so that merged modifications are not repeated in future operations. With n items in total and m modifications in a selected branch the expected cost is O(m log(n/m)).

## Acknowledgements

I would like to thank Niklas Röjemo for contributing to this paper with ideas and comments.## References

1. Eiichi Goto. Monocopy and associative algorithms in extended Lisp. Technical Report TR 74–03, University of Tokyo, May 1974.
2. J. Driscoll, N. Sarnak, D. Sleator, and R. Tarjan. Making data structures persistent. *J. of Computer and System Science*, 38:86-124, 1989.
3. C. Okasaki. *Purely Functional Data Structures.* Cambridge University Press, 1998.
4. J. Discroll, D. Sleator, and R. Tarjan. Fully persistent lists with catenation. *J. ASM*, 41(5):943-959, 1994.
5. A. Buchbaum and R. Tarjan. Confluently persistent deques via data structural bootstrapping. In *J. of Algorithms*, 18:513-547, 1995.
6. E. Demaine, S. Langerman, and E. Price. Confluently persistent tries for efficient version control. *Algorithmica*, 57(3):462-483, 2010.
7. A. Fiat and H. Kaplan. Making data structures confluently persistent. *J. Algorithms*, 48(1):16-58, 2003.
8. S. Collette, J. Iacono, and S. Langerman. Confluent Persistence Revisited. In SODA 2012: *Proceedings of the Twenty-Third Annual ACM-SIAM Symposium on Discrete Algorithms*, pages 593-601, Tokyo, Japan, 2012.
9. R. Seidel and C. R. Aragon. Randomized search trees. *Algorithmica*, 16:464-497, 1996.
10. G. E. Blelloch and M. Reid-Miller. Fast set operations using treaps. *ACM Symposium on Parallel Algorithms and Architectures - SPAA*, pp. 16-26, 1998.
11. G. Castagnoli, S. Braeuer, and M. Herrman. Optimization of cyclic redundancy-check codes with 24 and 32 parity bits. *IEEE Transact. On Communications*, Vol. 41, No. 6, June 1993.